\documentclass{article}
\font\mybold=cmmib10
\chardef\Myxi="18
\def\boldxi{\hbox{\mybold\Myxi}}
\newcommand{\mb}[1]{\mbox{\boldmath $#1$}}
\font\twelvemsb=msbm10 scaled 1000
\def\OskliveR{\hbox {\twelvemsb{R}}}
\date{}
\begin{document}

\title{Exact solutions and their interpretation} 

\author{JI\v{R}\'{I} BI\v C\'AK\\
{\small Institute of Theoretical Physics, Faculty of Math and Physics, Charles University}\\
{\small V Hole\v{s}ovi\v{c}k\'ach 2, 180 00 Prague 8, Czech Republic}\\
E-mail: \small \tt bicak@mbox.troja.mff.cuni.cz
}
\maketitle

\abstract{
This is the account of the workshop
{\it Exact solutions and their interpretation}
at the $16^{\rm th}$ International Conference on General
Relativity and Gravitation held in Durban, July 15-21, 2001. Work
reported in 32 oral contributions spanned a wide variety of topics,
ranging from exact radiative spacetimes to cosmological solutions.
Two invited review talks, on the role of exact solutions in string
theory and in cosmology, are also described.
}

\section{Introduction}
In accord with past tradition, the workshop A.1 has once again
received the highest number of papers (73) in GR16, thus providing
ample proof that the interest in discovering and understanding
solutions to the Einstein equations did not fade away with the
turn of the millennium. In comparison, a closely related workshop on mathematical
studies of the field equations received 48, three workshops
devoted to quantum issues totalled 52, mathematical cosmology
received 38 and relativistic astrophysics 31, etc.
Obviously, quantity does not necessarily mean quality. Also, not
all the contributors could have been expected to come to Durban.
Despite this, I asked the organizers to allot extra 3 hours to
the A.1 workshop, with the welcome side effect that no need of a
poster session arose. During 9 hours, 32 original
contributions (15 minutes each) were presented.
In addition, I turned to Roberto Emparan and Malcolm MacCallum to
give reviews (30 minutes each) on exact solutions in string theories
and on cosmological models. Their reviews also attracted a great
number of GR16 participants, who normally would not have attended talks
at the A.1 workshop.

In the Introductory remarks,  I very briefly recalled the role
of exact -- sufficiently explicit -- solutions for the development and
understanding of general relativity. Despite significant recent
advances in demonstrating the {\it existence} of some quite
general classes of solutions which, of course, are ``no less
exact'', we still rightly welcome  special {\it explicit}
formulae. Such questions as location and character of singularities and the
global properties of a corresponding
spacetime, are best addressed if an explicit solution is available.
A special explicit solution frequently stimulates questions
relevant to more general situations. For example, the presence of
the Cauchy horizon in the Reissner-Nordstr\"om solution inspired a
great amount of work on the instabilities of the Cauchy horizons
of other types.

The role of the most important exact vacuum solutions such as
black-hole solutions, stationary axisymmetric fields, and several
classes of exact radiative spacetimes and vacuum cosmological
models, in the development of general relativity and
relativistic astrophysics, was reviewed recently \cite{JeFest}.

In the following, the contributions are arranged in essentially
the same order as they were given during the workshop. Although
the topics covered were quite diverse, wherever possible, the
contributions were grouped together: on exact
radiative spacetimes, on the mathematical techniques used in
analyzing exact solutions, on cosmological solutions, on the
black-hole and other stationary solutions, and three contributions
are covered by ``Miscellanea''. The two invited reviews are
summarized at the end of the article.

With minor exceptions the following text is the combination of the
material written by those who gave the talks (with their names
in italics), which I have edited, modified in a number
of cases, and in all cases had to shorten
because of length constraints. I thank all the  authors for
supplying me with this material and, in particular, for adding the
references which make the report more useful. I also thank
T.~Ledvinka and M.~Varadarajan for help.

\section{Exact radiative spacetimes}

\textbf{Standing gravitational waves} ({\it H. Stephani})
\\
H. Stephani proposed an intuitive definition of standing gravitational
waves:
(i) the constitutive parts of the metric functions should depend on the
timelike coordinate only through a periodic factor, and they should also
depend on spacelike coordinates ("product structure"); (ii) the time
average of some of the metric functions should vanish; in particular,
the analogue of the Poynting vector (if there is any) should be
divergence-free and the time average of its spatial components should be
zero.
He then searched for standing gravitational waves among
the main classes of exact vacuum solutions -- in most cases with a negative
result. There are none among the plane waves and the algebraically special
solutions. He can find them only as simple subcases of the
(generalized) Einstein-Rosen waves. But in particular for the
Gowdy universes, the criterion ''product structure'' gives contradicting
results, depending on the choice of the coordinate system. The criterion
''Poynting vector'' again turns out to be ambiguous. The most probable
choice is that connected with the so-called
"C-energy" since the energy-momentum pseudotensor gives unconvincing
answers, and for the other forms based on a Lagrangian the timelike
component of the Poynting vector turns out to be negative.
Although Stephani could find Einstein-Rosen waves which meet his two
conditions, the coordinate dependence of the criteria remains disturbing.
One may expect more standing wave solutions in the classes with only one
or none spacelike Killing vector.
\\
\\\textbf{Cylindrical spacetimes endowed with angular momentum}\hfill\\
({\it G.~A.~Mena Marug\'an})\\
In this work the general solutions to the vacuum Einstein
equations with cylindrical symmetry were studied with the symmetry
axis being allowed to be singular -- it could be endowed with spin
and a linear energy density. First a gauge-fixing procedure
adapted to cylindrical symmetry that removes all the constraints
of canonical gravity was given. The corresponding reduced system
has two field-like degrees of freedom which depend only on the
radial and time coordinate. The gauge-fixed metric is expressed in
terms of these fields. The solution contains also two constant
parameters that describe the deficit angle outside the axis and
the density of the angular momentum in the axis direction. The
symmetry axis is regular if and only if the two constants vanish.
It is proved that the gauge-fixed metric is rigorously defined and
there exists a Hamiltonian that generates the reduced evolution if
one imposes suitable boundary conditions at spatial infinity and
on the axis on the two metric fields and, in addition, introduces
a certain restriction on the initial data if the angular momentum
does not vanish. The explicit form of the reduced Hamiltonian
which provides the energy density in the axis direction is also
found. This density is preserved in the evolution and turns out to
be bounded both from below and above. The considered spacetimes
describe cylindrical gravitational waves surrounding a spinning
string. Moreover, the deficit angle detected at spatial infinity
is equal or greater than the deficit seen close to the axis.
Gravitational waves have thus a positive contribution to the
energy density. For details, see Ref. [\cite{Marugan01}].
\\
\\\textbf{On Killing vectors and Bondi expansions} ({\it J.~A.~Valiente Kroon})\\
Asymptotically flat radiative spacetimes have frequently been
studied by solving a characteristic initial value problem by means
of expansions in inverse powers of a radial parameter (Bondi
expansions). If one assumes that the spacetime has a Killing
vector field, then by performing similar expansions on the Killing
equations it is possible to deduce the asymptotic form of the
Killing vectors compatible with asymptotic flatness.\cite{Kroon01}
Furthermore, one can deduce so-called ``constraint equations'' for
the news function, the mass aspect and the Newman-Penrose
constants. The particular form of these quantities is thus
restrained. By resorting to the analysis of the orbits of the
Killing vectors at null infinity, it is possible to give the
solution to the constraint equations. For Killing fields that are
non-supertranslational the characteristics of the constraint
equations are the orbits of the restriction of the Killing field
to null infinity. The possible classes of Killing fields are
discussed by analysing their orbits at null infinity. If axial
symmetry is imposed, and the spacetime is assumed to be radiative,
then one recovers known results on boost-rotation symmetry
as the only other admissible symmetry (see Ref. [\cite{Bicak01}] and
references therein). As an application, the formalism was used to
study the Einstein-Maxwell boost-rotation symmetric Petrov type $D$
spacetimes.\cite{Kroon02} Valiente Kroon also studied the late
time behaviour of the polarisation states, of the Newman-Penrose
constants and the Bondi mass. In particular, one finds that the
leading term behaviour of the Bondi mass loss is proportional to
the square of the norm of the only non-vanishing Newman-Penrose
constant. Since it is of quadrupolar nature, this result
resembles, at least formally, Einstein's quadrupole formula.
\\
\\\textbf{Boost-rotation symmetric radiative spacetimes
and fields: post-GR15 developments} ({\it J. Bi\v c\'ak}, P. Krtou\v s, V. Pravda and A.
Pravdov\'a)\\
These spacetimes, representing uniformly accelerated objects (e.g.
black holes), are significant since in rotationally symmetric,
locally asymptotically  flat at null infinity electrovacuum
spacetimes the only additional allowable symmetry that does not
exclude radiation is the boost symmetry (for latest formulations
and proofs, see Refs. [\cite{Bicak01,Kroon01}] and references
therein.) These solutions are the only explicitly known spacetimes
in which strong initial data can be chosen on a hyperboloidal
initial hypersurface which lead to complete, asymptotically
radiative spacetimes in future. Their properties and their use as
test-beds in approximation methods, numerical relativity and
quantum gravity models have been reviewed recently.\cite{JeFest,Bicak03}

The spinning C-metric is the only known example of such spacetime
with Killing vectors not hypersurface orthogonal. It was brought into the
canonical form of the boost-rotation symmetric spacetimes, and its interpretation
as a metric representing uniformly accelerated, rotating black holes was
analyzed.\cite{Bicak04} The geodesics in the standard C-metric
were also studied.\cite{Bicak_PC}

Most recently, scalar and electromagnetic fields produced by
uniformly accelerated (test) charges in de Sitter space were
analyzed.\cite{Bicak06} New effects arise because de Sitter space
has spacelike conformal infinities. The radiative properties of
the fields depend on the way in which a given point of {$\cal J$}
is approached. With particle horizons, purely retarded fields
become singular or even cannot be constructed at the "creation
light cone" of the "point" at which a source "enters" the
universe. Smooth fields involving both retarded and advanced
effects were constructed. In particular, the classical Born's
solution for the field of two uniformly accelerated charges in
Minkowski space were generalized to the de Sitter
universe.\cite{Bicak06}
\\
\\\textbf{Geodesics in expanding impulsive gravitational waves}
(J.~Podolsk\'y and {\it R.~Steinbauer})\\
Introduced in a classical paper of Penrose by using his ``scissors
and paste'' method, the explicit solution for spherical impulsive
gravitational waves in continuous coordinates was later on given
by Penrose and Nutku, and Hogan. It was only recently related
explicitly to the impulsive limit of the Robinson-Trautman type
$N$ solutions.\cite{Steinbauer_gp} The latter however has to be
considered as formal since the metric tensor contains terms
proportional to the square of $\delta$-function. The
transformation relating the distributional coordinate system to
the continuous one is discontinuous. It is analogous to the one
relating the distributional to the continuous form of the metric
tensor for impulsive pp-waves which has been analyzed rigorously.\cite{Steinbauer_ks}

The authors investigated the geodesics in spacetimes with spherical impulsive waves.
In particular, employing the continuous form of the metric they found a large class of privileged
and simple geodesics which can be related to the geodesics in the distributional form of the metric
``in front'' and ``behind'' the spherical impulse. This provides foundations for a
rigorous (distributional) treatment of impulsive Robinson-Trautman solutions of type $N$.

\section{Mathematical methods}

\textbf{Killing vectors, homothetic vectors, and conformal Killing vectors in the Newman-Penrose formalism}
({\it G. Ludwig} and B. Edgar)\\
Finding Killing vectors (KV) or a homothetic vector (HV) of a
given metric can be a formidable task. The authors present a new
and simpler approach within the Newman-Penrose formalism. Leaning
on results\cite{Ludwig01,Ludwig02} previously derived in the
Geroch-Held-Penrose formalism they show that the Killing or
homothetic equations can be replaced by the NP-Lie equations, a
set of equations involving the commutators of the Lie derivative
with the four NP differential operators applied to the four
coordinates. Provided that these operators refer to a preferred
tetrad relative to the KV or HV, the equations can be solved for
the Lie derivative of the coordinates, i.e., for the components of
the KV or HV. If part of the tetrad (null directions and gauge)
can be defined intrinsically, then that part is generally
preferred relative to any KV or HV or conformal Killing vector. In
case when part of the tetrad cannot be defined intrinsically, the
tetrad used is defined only up to one null rotation parameter
and/or a gauge factor, and the NP-Lie equations become more
involved. The general method remains the same and is still more
efficient than conventional methods. See Ref. [\cite{Ludwig03}]
for details. The following contribution is devoted to a closely
related problem.
\\
\\\textbf{Tetrads and symmetries}
({\it B. Edgar} and G. Ludwig)\\
Two of the main tools in the search for exact solutions of
Einstein's equations are tetrad formalisms and use of symmetry.
The authors propose a method which  analyzes efficiently the
Killing vector structure of metrics as they are calculated in
tetrad formalisms; this method builds  on the following
results:\cite{Edgar02} a vector field $\boldxi$ is a Killing
vector field iff there exists a  tetrad $\hat {Z}_m^\mu $ ($
m=1,2,3,4$), which is Lie derived with respect to this field,
${{\mathcal L}}_\xi (\hat {Z}_m^\mu)=0$. If we are  given, in some
$ Z_m{}^{\mu}$, a spacetime containing  at least one  KV
$\boldxi$, there exists a  tetrad $\hat {Z}_m{}^{\mu}$ which is
Lie derived by that KV and which  is given by a transformation
linking $\hat {Z}_m{}^{\mu}$ to $ {Z}_m{}^{\mu}$ for some values
of the six Lorentz parameters. A direct procedure to find KVs is to
substitute $ Z_m{}^{\mu}$ via $\hat {Z}_m{}^{\mu}$ and six
arbitrary parameters into ${{\mathcal L}}_\xi (\hat {Z}_m{}^{\mu})
= 0$, and integrate this system for $\boldxi$ and for all the
parameters simultaneously. The explicit KVs and the parameters
giving $ \hat Z_m{}^{\mu}$ are then obtained. Rather than
integrate  ${{\mathcal L}}_\xi (\hat {Z}_m{}^{\mu})=0$, it is
efficient\cite{Ludwig01} to deal with the equivalent commutator
equations for ${{\mathcal L}}_\xi $ and $\hat \nabla_m = \hat
Z_m{}^{\mu}\nabla_\mu $.
\\
\\
\textbf{Conformally decomposable spacetimes} ({\it J.$\;$Carot} and B.$\;$O.$\;$J.$\;$Tupper)
\\
Here spacetime $(M, g)$ is studied whose metric is conformally
related to the metric of a 2+2 locally decomposable spacetime
$(M,\tilde{g})$; i.e., it is a product of two 2-dimensional
spaces. Starting from the fact that the underlying decomposable
spacetime cannot posses proper conformal Killing vectors (CKV)
(unless it is conformally flat),\cite{Carot_ColeyTupper} the authors
show that the CKV of $(M, g)$ must be the Killing and
homothetic vector fields of $(M,\tilde{g})$ and carry out a
complete classification of all the different possibilities for the
conformal Lie algebra of $(M, g)$ deriving canonical forms for
both the metric and the CKV. A number of results on KV and HV
fields are reviewed and new ones derived, especially those
concerning homotheties and isometries with fixed points. The
authors also obtain some classes of physically relevant examples,
such as perfect fluids and electromagnetic fields.
\\
\\\textbf{Equations for complex-valued, twisting, type-{\it N} vacuum solutions,
with one or two Killing/homothetic vectors}
({\it D. Finley})\\
The goal of understanding solutions of Petrov type $N$ with
non-zero twist is still not realized. Although they can hardly
represent radiative fields of bounded
sources,\cite{Finley_BicakWei} they attract researches because of
a rich mathematical structure. One way to understand them is the use
of HH spaces, i.e., complex-valued spacetimes, of Petrov type
$N\!\times N$. They are determined by three PDEs for two
functions, $\lambda$ and $a$, of three independent variables (and
also two gauge functions which may be chosen to be two of the
independent variables). As in integrable systems, these form a
second order, linear system for $\lambda$; however, here the
integrability conditions, involving $a$, are complicated.
Therefore, with the hope of finding new solutions, Finley assumed
that these spacetimes admit both one and two homothetic or Killing
vectors.  The case with one Killing and one homothetic vector
reduces equations to two ODEs for two unknown functions of the one
remaining variable. Finley described the explicit forms of the
metric, tetrad, connections, and curvature for twisting HH spaces
of Petrov type $N\!\times N$, modulo the determining equations.
The details can be found in recent Ref. [\cite{Finley_Fin}].
\\
\\\textbf{Algebraic approach to the study of spacetimes with an isometry} ({\it C. F. Sopuerta} and F. Fayos)
\\
In this contribution a new formalism for the study of spacetimes
possessing a non-null Killing vector $\mb{\xi}$ is presented. The
main quantity in this approach is 2-form $\mb{F} = \mb{d\xi}\,$
satisfying Maxwell's equations. Considering
$(\xi^\alpha,F_{\alpha\beta})$, the equations $\nabla_\alpha
\xi_\beta = \textstyle{1\over2}F_{\alpha\beta}$,
$\nabla_{[\alpha}F_{\beta\gamma]}=0$,
$\nabla_{\beta}F^{\alpha\beta}=J^\alpha=2R^\alpha{}_\beta\xi^\beta$,
and studying the integrability conditions, the authors find that
all the components of the Weyl tensor can be expressed in terms of
$(g_{\alpha\beta},\xi^\alpha,F_{\alpha\beta},T_{\alpha\beta})$,
$T_{\alpha\beta}$ being the energy-momentum tensor. An extension
of the Newman-Penrose formalism is set up in which a null tetrad
adapted to $F_{\alpha\beta}$  is chosen, and the tetrad components
of $\xi^\alpha$ and $F_{\alpha\beta}$ are used. The dependence of
$C_{\alpha\beta\gamma\delta}$ on
$(g_{\alpha\beta},\xi^\alpha,F_{\alpha\beta},T_{\alpha\beta})$ is
algebraic.  Substituting $C_{\alpha\beta\gamma\delta}$ into the
second Bianchi identities one obtains first-order equations for
the spin coefficients.  Therefore, all the compatibility
conditions reduce to compatibility conditions for the first-order
equations for the spin coefficients. This formalism naturally
provides a refinement of the Petrov classification: apart from the
algebraic structure of $C_{\alpha\beta\gamma\delta}$ one has the structure of $F_{\alpha\beta}$ (see Ref. [\cite{Sopuerta_FS1}]).
\\
\\\textbf{Painlev\'e analysis in general relativity} ({\it R. Halburd})
\\
This talk illustrated the use of complex-analytic methods as
``detectors'' of solvable models. The problem of determining the
metric for a non-static shear-free spherically symmetric fluid
(charged or neutral) reduces to the problem of determining a
one-parameter family of solutions to
${d^2y}/{dx^2}=f(x)y^2+g(x)y^3$, where $f$ and $g$ are arbitrary.
Most of the solutions are special cases of the family given by
Sussman. It was shown\cite{halburd} that Sussman's solutions
correspond to special cases such that the general solution has a
simple singularity structure in the complex domain (the Painlev\'e
property). New solutions were also given in terms of Airy
functions and Painlev\'e transcendents. Halburd described all
choices for $f$ and $g$ for which the equation above admits a
one-parameter family of solutions such that all movable
singularities are poles. This is weaker than the
Painlev\'e property and yields a number of solutions described by
Riccati equations besides those given in Ref. [\cite{halburd}].
Conditions for matching these to an external
metric were determined.

\section{Cosmological Solutions}

\textbf{Hypersurface homogeneous rotating dust models} ({\it A. Krasi\'nski})
\\
For matter that moves geodesically and with nonzero rotation,
Pleba\'nski coordinates (P.c.) $(t, x, y, z)$ can be chosen so
that the velocity field, the metric and the rotation vector field
have the forms $ u^{\alpha} = \delta^{\alpha}_0$, $g_{0 \alpha} =
\delta^0_{\alpha} + y \delta^1_{\alpha}$, $w^{\alpha} =
n\delta^{\alpha}_3$, where $n$ is the number density of the
particles of the matter. If any symmetries exist, then every
Killing field, expressed in P.c., must have the form $ k^{\alpha}
= (C - \phi - y \phi,_y)\delta^{\alpha}_0 +
\phi,_y\delta^{\alpha}_1 - \phi,_x \delta^{\alpha}_2 + \lambda
\delta^{\alpha}_3, $ where $\phi(x, y)$ and $\lambda(x, y)$ are
arbitrary functions and $C$ constant. If $\phi = $ constant, then
$k^{\alpha} = Cu^{\alpha} + \lambda w^{\alpha}$, and this property
is invariant under all the allowed transformations of the P.c. If
$\phi \neq $ constant, coordinates can be adapted to this Killing
field so that $k^{\alpha} = \delta^{\alpha}_1$. When 3 Killing
fields exist, a classification of all the metric forms can be
provided. The corresponding symmetry algebras can all be assigned
to the Bianchi classes; the orbits can  be spacelike as well as
timelike or null. Apart from a few classes in which the orbit has
to be timelike, most of the 25 classes contain a free parameter
that measures the tilt of the velocity field with respect to the
symmetry orbits, and allows one to consider all possible positions
of the orbits relative to velocity, including the nonrotating,
isotropic and spatially homogeneous Friedmann limit. The
results have all been published by A. Krasi\'nski in J. Math.
Physics (see Ref. [\cite{Kasinski01}] and citations therein).
\\
\\\textbf{Spherical inhomogeneous cosmologies with pressure} ({\it J. R. Gair})
\\
A generalization to the Lema\^{\i}tre-Tolman-Bondi solution may be
obtained describing a spherically symmetric universe filled with
dust which has angular momentum. The dust is labeled by a comoving
radial coordinate, $R$, such that each dust particle remains on a
surface $R =$ constant. The dust particles within each of these
shells move in every tangential direction with a certain angular
momentum about the symmetry centre. This angular momentum is
conserved and so it is a function of shell label only. It produces
an effective tangential pressure, the radial pressure vanishes.
Einstein derived a static solution of this form, Datta and Bondi
generalized to the dynamic case, but did not obtain the metric,
Magli derived the metric in mass-area coordinates. Now Gair has
obtained a new solution\cite{Gair01} in terms of more physical
coordinates - the shell label, $R$, and the proper time, $\tau$,
experienced by the dust particles. The equation of motion for the
areal radius, $r$, is of the form $(\partial r/\partial \tau)
=\sqrt{E(R)-V(r,R)}$. The potential $V(r,R)$ for a given shell is
the potential of a test particle moving in a Schwarzschild-de
Sitter metric. The solution contains four arbitrary functions of
$R$ -- the angular momentum $L(R)$, the mass $M(R)$, the `energy'
$E(R)$ and the initial position of the shells, $\tau_{0}(R)$. The
shape of the potential permits seven different types of evolution
for each shell: expansion and recollapse, collapse and bounce,
unbounded expansion/collapse, oscillations, static, coasting
expansion/collapse and hesitation. Self-similar solutions may be
obtained by choosing $M \propto R$, $L \propto R$ and $E =$
constant, and a solution with a null fluid source by letting $L
\rightarrow \infty$.
\\
\\\textbf{On some perfect fluid solutions of Stephani} ({\it A. Barnes})
\\
In 1987 H. Stephani derived several solutions for a geodesic
perfect fluid flow with zero pressure (dust). Barnes
generalized\cite{Barnes01} these solutions with non-zero rotation
to include a non-zero cosmological constant $\Lambda$. All
solutions are of Petrov type $D$ and the magnetic part of the Weyl
tensor (relative to the fluid flow) vanishes. In general the
solutions admit no Killing vectors. The fluid flow is shearing,
twisting and expanding. In the limit when the energy density
vanishes the solutions reduce to spacetimes of constant curvature.
The solutions can be matched across timelike hypersurface to a de
Sitter, anti de Sitter or Minkowski spacetime, depending on $\Lambda$.
\\
\\\textbf{Locally isotropic spacetimes} (M. A. H. MacCallum and {\it F. C. Mena})
\\
The authors consider spacetimes which locally (at every point in a
neighbourhood) have a non-trivial isotropy group. The possible
cases are defined in terms of symmetry of the Riemann tensor and
its derivatives, following the method introduced by Cartan and
Karlhede (see Ref. [\cite{Mena01}]) which gives a unique local
characterization of a spacetime. They review and sharpen previous
results for the cases with continuous isotropy groups. For the
discrete case they analyze the possible maximal isotropy groups
allowed by the algebraic structure of the Riemann tensor. The aim
is to prove results analogous to those of Schmidt from 1969, who
showed that a perfect fluid spacetime which locally admits the
reflections in three perpendicular spatial axes as isotropies must
admit a 3-parameter group of isometries. The authors obtain
several results of this sort for groups of spatial reflections.
For example, they show that for a spacetime with a local discrete
isotropy group $L$ generated by reflection of a plane, whose
energy-momentum tensor is that of a vacuum or perfect or imperfect
fluid, if $R_{\alpha\beta\gamma\delta}$,
$R_{\alpha\beta\gamma\delta;\mu\nu}$ and
$R_{\alpha\beta\gamma\delta;\mu\nu\rho}$ are invariant under $L$,
then there is a local group $G_2$ of isometries acting
transitively on the surfaces to which the reflection plane is
tangent.
\\
\\\textbf{On stationary, spherically symmetric inhomogeneous spacetime with matter content
obeying $p=\alpha \rho$} ({\it S. M. Wagh}, D. W. Deshkar and P. S. Muktibodh)
\\
The authors consider the shear-free, spherically symmetric,
inhomogeneous spacetimes admitting the separable metric of the
form\cite{Wagh_cqg1}
$$
ds^2 = -y^2A^2dt^2 + 2(y')^2R^2dr^2 +
y^2R^2\left( d\theta^2 + \sin^2{\theta}\,d\phi^2\right)
$$
where $y=y(r)$, $A=A(t)$, $R=R(t)$. The coordinates are comoving
and the spacetime has non-vanishing energy flux. One nontrivial
field equation and the equation of state determine $R(t)$. For
example, the imperfect matter may obey $p=\alpha\rho$, $\alpha=$
constant. The radial distribution of $\rho$ is unspecified since
the field equations do not determine the metric function $y(r)$
and $\rho \propto 1/y^2$. A timelike Killing vector exists when
${\dot{R}}/{A} = {\rm constant}$. The stationary solution has an
application for an inhomogeneous steady state cosmology. In the
absence of a timelike KV, the metric can describe an
inhomogeneous big bang cosmology.

\section{Stationary systems and black holes}

\textbf{On electromagnetic Thirring problems} (M. King and {\it H. Pfister})
\\
The authors consider models consisting of two spherical shells of
radii $a$ and $R\ge a$, the first one carrying a charge $q$ but no
rest mass, the second one mass $M$ but zero charge (the mass shell
fulfils the weak energy conditions). To these shells small angular
velocities are applied. The coupled Einstein--Maxwell equations
were solved exactly in $M/R$ and $q/R$, and to first order in the
angular velocities. The dragging effects and the induced magnetic
fields were calculated.\cite{King01} The authors confirm some
results of Cohen, and Ehlers and Rindler; in contrast to the
latter they argue that the results completely fulfil Machian
expectations. The collapse limit of the system was also discussed.

Then only one slowly rotating shell carrying both mass and charge
was considered. For $M/R$ and $q/R$ for which the weak energy
condition is violated, the dragging constant can become negative
(antidragging!). Interesting is the behavior of the gyromagnetic
ratio: it is very near to $G=2$ in most of the parameter space,
e.g., $1.92<G<2.05$ for all $M/2R>1$ and $q/R$ that the weak
energy condition is fulfilled. The authors argue that the
``robustness'' of the value $G=2$ and its ``coincidence'' with
$G\approx 2$ for the simplest rotating, charged particles (electron and muon) hints
to a deep connection between general relativity and quantum theory.
\\
\\\textbf{Exact relativistic treatment of stationary counterrotating dust disks}
(J.~Frauendiener and {\it C.~Klein})
\\
Relativistic dust disks with counter-rotating matter have been discussed
since the work of Morgan and Morgan. The rigidly rotating single
component dust disk was treated numerically and perturbatively by
Bardeen and Wagoner and solved analytically by Neugebauer and
Meinel in terms of Korotkin's finite gap solutions. The authors
discuss a class of stationary counter-rotating dust disks which
interpolates continuously between the rigidly rotating disk and a
static disk. The whole metric is given explicitly in terms of
theta functions on a hyperelliptic Riemann surface. The metric
functions are analyzed analytically and the hyperelliptic
functions evaluated numerically by using spectral methods.
Physically interesting limiting cases as the Newtonian, the
static, and the ultrarelativistic limit, where the central
redshift diverges, are studied. Explicit expressions for the mass
and the angular momentum are given. Following Bi\v c\'ak and
Ledvinka who studied counter-rotating disks of infinite extension
as sources for the Kerr metric the authors show that the matter in
the disk can be interpreted as two streams of pressureless matter
which move on geodesics of the inner geometry of the disk. The
occurence of ergospheres and the dragging of the inertial frames
is also discussed. See Ref. [\cite{Klein08}] for details and references.
\\
\\\textbf{Generalized Kerr-Schild metrics and the gravitational field of a
massless particle on the horizon} ({\it H. Balasin})
\\
Following the work of Aichelburg and Sexl, and \( \mbox {'t\,
Hooft} \) and Dray on the gravitational field of a massless
particle in Minkowski space, and on the horizon of a Schwarzschild
black hole, respectively, Balasin investigated the change in the
spacetime structure generated by a massless particle moving along
the horizon \( \mathcal {H} \) of an arbitrary stationary black
hole. The generalized Kerr-Schild class characterized by
$\tilde{g}_{\alpha\beta}=g_{\alpha\beta}+fk_{\alpha}k_{\beta}$,
$k^{\alpha}k^{\beta}g_{\alpha\beta}=k^{\alpha}k^{\beta}\tilde{g}_{\alpha\beta}=0$,
$(k\nabla )k^{\alpha}=(k\tilde{\nabla})k^{\alpha}=0$ gives a
framework well adapted to the problem. Identifying the null
generators of the horizon with the geodetic null vector field \(
k^{a} \) and taking the scalar function \( f \) to be concentrated
on the horizon, i.e. \( f=\delta _{\mathcal {H}}\tilde{f} \),
allows one to investigate the situation in terms of properties of
supporting null hypersurface (the horizon) and the parent geometry
represented by \( g_{\alpha\beta} \). Balasin obtained a
generalized \( \mbox {'t\, Hooft} \)-Dray equation for the reduced
{}``profile'' \( \tilde{f} \), whose coefficients are determined
by optical scalars of the congruence forming \( \mathcal {H} \)
and curvature quantities thereon. See Ref. [\cite{Balasin01}] and
references therein.
\\
\\\textbf{The generalized Wahlquist-Newman solution}({\it M. Mars})
\\
The Kerr metric and the Wahlquist metric share the property that
their Simon tensor vanishes. Recent work has clarified the
geometrical meaning of the vanishing of the Simon tensor in terms
of a natural relationship between the Weyl tensor and the
covariant derivative of the stationary Killing vector (the
so-called Papapetrou form). The Kerr-Newman-de Sitter spacetime
satisfies the same relationship.\cite{Mars01} The electromagnetic
field and the Papapetrou form are also closely connected in this
metric. This leads to the likely existence of a charged
generalization of the Wahlquist metric -- Wahlquist-Newman metric.
This was explicitly found in Ref. [\cite{Mars01}]. It was
originally given by Garc\'{\i}a using different hypotheses. This
family contains eight essential parameters and includes, besides
the Wahlquist and Kerr-Newman-de Sitter metrics, the whole
Pleba\'nski limit of the rotating C-metric. A weakening of the
relationship between the Weyl tensor and the Papapetrou form leads
to a larger family containing one arbitrary function of one
coordinate. The family is generically of Petrov type $I\!I$ and
possesses only one Killing vector. These metrics admit a
well-defined ``infinity'', with topology $\OskliveR \times S $,
where $S$ is a two-dimensional manifold, which may be compact or
not. The metrics are in general not asymptotically flat but have a
quite rich asymptotic behaviour.
\\
\\\textbf{General vacuum solution for axially symmetric stationary spacetime}
(Z. Ya. Turakulov and {\it N. Dadhich})
\\
The authors have obtained the most general vacuum solution for the
axially symmetric stationary metric in which both the
Hamilton-Jacobi equation for particle motion and the Klein-Gordon
equation are separable. It can be transformed
into the Kerr-NUT solution. Like the Kerr solution, the Kerr-NUT
solution is thus also unique under these assumptions. It is the
most general axially symmetric stationary asymptotically non-flat
vacuum spacetime admitting regular horizon. See Ref. [\cite{Dadhich01}] for details.
\\
\\\textbf{On the integrability of the exact and perturbative Einstein's equations for axistationary perfect fluids}
({\it M.~Bradley}, G.~Fodor, M.~Marklund and Z.~Perj\'es)
\\
In this contribution the integrability of stationary axisymmetric
perfect fluid spacetimes was investigated using a tetrad approach
with the Riemann tensor, the Ricci rotation coefficients and the
tetrad vector components as
variables.\cite{bradley_bm,bradley_fmp} The system of equations is
integrable in the generic case according to the Cauchy-Kowalewski
theorem. The equation of state is barotropic if and only if
$\omega_1\sigma_2=\omega_2\sigma_1$ holds, where $\omega_i$ is the
vorticity and $\sigma_i$ the shear (with tetrad $e_1$ and $e_2$
in the "$r,\theta$"-plane). A physically realistic rotating
incompressible perfect fluid must be of Petrov type
$I$. If the equation of state is linear, the
Petrov type cannot be $D$. The authors linearize the equations
around the interior Schwarzschild solution. The resulting solution
of the linearized equations is known to be a first order
approximation of a solution to the full non-linear system because
of the integrability of the system. An example of the integration
was given.
\\
\\\textbf{Plane symmetric analogue of NUT space} ({\it M.~Nouri-Zonoz} and A.~R.~Tavanfar)
\\
The authors introduce a new definition of spacetime symmetry which
is in accordance with the symmetry of its curvature invariants and
its gravoelectromagnetic fields. Applying this definition, they
investigate exact vacuum solutions corresponding to both static
and stationary {plane symmetric} spacetimes using concepts of the
(1+3)--decomposition or {threading} formalism. Demanding the
presence of a plane symmetric gravomagnetic field, they find a
family of  two parameter ($m$ and $l$) solutions, every member of
which is the plane analogue of NUT space. See Ref.
[\cite{Nouri01}] for details.
\\
\\
\textbf{New exact solution to the static balance problem} ({\it R. B. Mann} and T. Ohta)
\\
The static balance problem of finding equilibrium solutions in
which the gravitational attraction of two bodies is balanced by
their electromagnetic repulsion was solved by Majumdar and
Papapetrou. These solutions are extremal in the sense that the
mass of each body is equal to its charge in gravitational units,
$e_{i}=\pm \sqrt{4\pi G}m_{i}$. No one has yet found equilibrium
states in which \ this condition does not hold for both bodies.
Within $(1+1)$ dimensional (``lineal'') gravity
the authors considered the motion of two charged gravitating
bodies. They solved for the Hamiltonian as a
function of the degrees of freedom of the system, which in this
case are the centre-of-inertia momentum and the relative proper
separation of the bodies. Demanding that the centre-of-inertia
momentum $p$\ had vanishing time derivative they found the
equation\cite{Mann01} $
{\kappa/2}\left( \sqrt{p^{2}+m_{1}^{2}}-p\right) \left( \sqrt{%
p^{2}+m_{2}^{2}}-p\right) -e_{1}e_{2}=0.
$
This condition is the generalization of the Majumdar-Papapetrou condition.
It depends on the momentum and can be
satisfied for some fixed momentum $p_{c}$ whilst $e_{i}\neq \pm \sqrt{4\pi G}m_{i}$.
\\
\\
\textbf{K\'ota-Perj\'es metrics and their Kerr-Schild class} ({\it M. Vas\'uth})
\\
In this work properties of Kerr-Schild metrics and K\'ota-Perj\'es
spacetimes were discussed. Gergely and Perj\'es have examined
under which conditions the Kerr-Schild pencil of a vacuum metric
generates another vacuum spacetime. Using parameter $\eta$, which
governs the shear of the null congruence, one can classify the
possible solutions of the problem. When $\eta$ is arbitrary, the
solution is either Kasner metric or another K\'ota-Perj\'es metric
depending on the curl of the null congruence. In the original
paper there are three K\'ota-Perj\'es metrics. In the above
classification only two of them were listed. The author shows that
the third metric is a mate of the other K\'ota-Perj\'es metric in
the class determined by the same special value, $\sin\eta=\pm
1/\sqrt{2}$, of the parameter. Some global properties of this
third metric were analyzed.
\\
\\
\textbf{De Sitter-Schwarzschild geometry} ({\it I. G. Dymnikova})
\\
Under spherical symmetry, the weak energy condition, the
regularity at the center, the asymptotic flatness and the
finiteness of the ADM mass define the family of regular solutions
which include the class of metrics asymptotically de Sitter as
$r\rightarrow 0$ and asymptotically Schwarzschild as $r\rightarrow
\infty$. The source is anisotropic perfect fluid which can be
interpreted as a vacuum invariant under boosts in the radial
direction.\cite{Dymnikova_2} According to the author, it connects
the de Sitter vacuum in the origin with the Minkowski vacuum at
infinity and corresponds to an extension of $\Lambda g_{\mu\nu}$
to the cosmological tensor $\Lambda_{\mu\nu}$ which allows
$\Lambda$ to become evolving and clustering.\cite{Dymnikova_4}

With masses $M>M_{crit}$, the spacetime contains an infinite
sequence of vacuum non-singular black and white holes whose
singularities are replaced with regular cores asymptotically de
Sitter as $r\rightarrow 0$. A white hole core can model initial
stages of the universe starting from nonsingular big
bang, followed by the Kasner-type expansion.
The author suggests that the instability of de Sitter vacuum near the
regular surfaces $r=0$ results in a possibility of multiple
quantum birth of baby universes inside.\cite{Dymnikova_4} A gravitating particle-like structure without horizons with
de Sitter vacuum in the origin was applied to estimate the lower
limits on sizes of fundamental particles.
\\
\\
\textbf{Hairy black holes, horizon mass, and solitons} ({\it D. Sudarsky}, A. Ashtekar and A. Corichi )
\\
In this work colored static black hole solutions to the
Einstein-Yang-Mills (EYM) equations and hairy black holes in other
theories were considered within the isolated horizons
framework.\cite{Wisniewski_let} Hairy black holes may be regarded
as `bound states' of ordinary black holes without hair and colored
solitons. This model is used to predict the qualitative behavior
of the horizon properties of hairy black holes, to provide a
physical `explanation' of their instability and to put qualitative
constraints on the end point configurations that result from this
instability. The model explains qualitative aspects of the
behavior of surface gravity of the colored  black holes.
Predictions for the value of the  ADM mass of the soliton
configurations found often in association with the hairy black
holes in, for example, Einstein-Yang-Mills-Higgs theory can be
made. These predictions allow one to evaluate the  difference in
the values of the mass of these solitons in terms of quantities
associated solely with the corresponding black holes and the
values of the ADM masses of the various Bartnik-McKinnon solitons.
The results have  been reported in Ref. [\cite{Sudar_acs}],
numerical calculations in Ref. [\cite{Sudar_num}].
\\
\\
\textbf{Isolated horizons in 2+1 dimensions}
\\({\it J. Wi\'sniewski}, A. Ashtekar and O. Dreyer)
\\
The framework of isolated horizons\cite{Wisniewski_let} is
here extended to $2$+$1$ dimensional general relativity with
negative cosmological constant. An analog of the
orthonormal null tetrad is introduced, where $l^{\mu}$, $n^{\nu}$
are null, $l^{\mu}$ is tangent to the horizon and $m^{\rho}$ is
spacelike. One can then construct the analog of the
Newman-Penrose framework.\cite{Wisniewski_tpo}
A (weakly) isolated horizon is a pair $(\Delta,[l])$, where
$\Delta$ is a null surface and two $l^\mu$'s are equivalent if
they differ by a multiplicative constant. Expansion of
$\Delta$ vanishes and the intrinsic connection 1-form on
$\Delta$ is Lie-dragged by $l^\mu$. Equations of motion are
imposed on $\Delta$ and an energy condition on stress-energy
tensor is assumed. The consequences of this definition are:
intrinsic metric on $\Delta$ is Lie-dragged by $l^\mu$, the $0$-th
law of black hole mechanics holds, and one can always choose a
gauge in which electrostatic potential is constant on $\Delta$.
Using Hamiltonian methods as in $3$+$1$, the quasi-local
definitions of mass and angular momentum can be introduced. The
first law of thermodynamics is equivalent to the requirement that
the time evolution vector field $t^{\mu}$ gives rise to a
Hamiltonian vector field on the phase-space of isolated horizons.
However, in presence of non-zero total charge, the appropriate
boundary conditions turn out to be more subtle than in $3$+$1$
dimensions. Energy can only be uniquely defined up to an additive
function of the electric charge.
The concepts introduced were used to analyze the charged
generalization of BTZ solution.
\\
\\
\textbf{Asymptotic behaviour of the proper length and volume of the Schwarzschild singularity}
({\it A. Qadir} and  A. A. Siddiqui)
\\
It has been proved that the proper length
of the Qadir-Wheeler suture model, constructed from two sections
of two closed FLRW models joined by a Schwarzschild region,
goes to infinity as the
singularity is approached, while its proper volume shrinks to zero.\cite{Qadir01} Asymptotically, the
proper length goes as $K^{1/3}$ and the proper volume as $K^{-2/3}$, $K$
is the mean extrinsic curvature. A similar analysis could
provide the asymptotic behaviour of the Schwarzschild black hole
near the singularity -- with $K$
taken as a time parameter. Trivial foliations, as by constant
Kruskal time hypersurfaces, or by hypersurfaces of
constant values of the compactified Kruskal time, do show
that the proper length approaches zero and the volume approaches infinity.
The relevant K-slicing for the Schwarzschild geometry is not so simple, the difference
from the suture model being that it has a boundary which limits the
part of the region near the singularity whose proper length is to
be measured. The authors prove that, as for the suture model, the
length does indeed go as $K^{1/3}$ and the volume as $K^{-2/3}$.

\section{Miscellanea}

\textbf{The problem of coordinates in general relativity} ({\it A. Chamorro})
\\
The author defines quasi-Minkowskian locally inertial coordinates
(QMLIC) relative to an observer in free fall that is
instantaneously at rest at an event by usual requirements: (i) the
world line of the observer (WLO) is described in the QMLIC by zero
spatial coordinates, time is observer's proper time; (ii) on the
WLO the metric coincides with Minkowski metric and its derivatives
vanish; (iii) the coordinate lines of the QMLIC have as unit
tangent vectors, an orthonormal tetrad which is parallelly
transported along the WLO, one vector being WLO's tangent; (iv) if
the tangent vectors have components which are the derivatives of
the original coordinates with respect to each of the QMLIC, then,
in a neighborhood of the event, their absolute differentials
differ from zero by a curvature dependent term. The author shows
that in a weak gravitational field there are in general infinitely
many different QMLICs all of which reduce to the unique
Minkowskian coordinates when the curvature vanishes. Once the
curvature term  and the tetrad are fixed, so are
the corresponding QMLIC.
\\
\\
\textbf{Coframe teleparallel models of gravity. Exact solutions} ({\it Y.~Itin})
\\
Itin studies the  coframe teleparallel theory of gravity with the most general
quadratic Lagrangian.\cite{Itin_hehl95} The coframe field on a differentiable manifold is
a basic dynamical variable. A metric tensor as well as a metric
compatible connection are generated by the coframe in a unique manner.
For a special choice this theory gives an alternative  description of
Einsteinian gravity - teleparallel equivalent of GR.
The field equations of the theory were studied by a ``diagonal'' coframe
Ansatz.\cite{Itin01} Itin obtained the explicit form of all spherically symmetric static
solutions of the ``diagonal'' type to the field equations for an
arbitrary choice  of free parameters and proved that the unique
asymptotically flat solution with Newtonian limit is the Schwarzschild
solution that holds for a subclass of teleparallel models.
\\
\\
\textbf{Matching preserving the symmetry} ({\it R. Vera})
\\
The matching of spacetimes is usually implicitly assumed to
preserve some of the symmetries of the problem. Some previous work
performed matchings by taking into account the orbits of the
preserved groups at both sides of the matching hypersurface. In
the work of Vera,\cite{Vera_raultesi} a general definition for
such a kind of matching is given. The matching hypersurface is
restricted to be `tangent' to the orbits of a desired group of
isometries admitted at both sides. The definition implies that the matching hypersurface
inherits the preserved symmetry and its algebraic type. When matching
is preserving a 2-dimensional group of isometries, two functions
constructed from the exterior product of the two Killing forms and
the differential of one of the Killing forms 
at one side, and the analogous two functions at the other side,
have to coincide at the matching hypersurface. The orbits of the
group generate orthogonal surfaces if and only if these two
functions vanish. This is related to the `circularity condition'
in stationary, axisymmetric interior problems. This implies the
absence of convective motions in fluids without flux. The
functions vanish in the vacuum exteriors and, therefore, have to
vanish also on the matching hypersurface.

\section{Invited reviews}

\textbf{The role of exact solutions in string theory} ({\it R. Emparan})
\\
There are at least two levels at which one can discuss classical exact
solutions within the context of string theory: (i) classical solutions
of full string theory, and (ii) classical solutions of string theory at
low energies. The latter are approximations to the former, where terms
in the equations that could give rise to corrections at curvature
scales of the order of the string length are neglected. Solutions in
(ii) are typically exact solutions of GR, possibly
coupled to a variety of scalar and p-form gauge fields. Only a
few among these belong to the more stringent class (i).

The techniques for constructing exact solutions to the full classical
string theory are rather different from those familiar in GR.
Conformal sigma models provide automatically solutions of this sort (see, e.g., Ref. [\cite{Emparan_wzw}]).
Besides, supersymmetry is often crucial in establishing the absence of
stringy corrections to some solutions of the low energy effective
action, such as the extremal Reissner-Nordstrom black holes, pp-waves,
extremal p-branes, AdS spacetimes etc.\cite{Emparan_stelle}   Supersymmetry is
also of help in solving the field equations, since it reduces them to
first order partial differential equations.

Many of the techniques developed for GR can be used in low energy string theory.
String theory provides additional
solution-generating techniques, based on the existence of dualities (S
and T) which relate solutions to different low energy limits of
string/M-theory. Other techniques make use of additional, compactified
dimensions: for example, performing boosts along them give rise to electric
Kaluza-Klein charges.

Exact gravitating solutions were far from the center stage in string
theory until around the late '80s. Since '96--'97 they have come to play
a crucial role, following the successful microscopic description of
black holes using D-branes,\cite{Emparan_stva} and the formulation of the
AdS/CFT correspondence, reviewed in Maldacena's
contribution to this conference. Besides this, some old and well-known
exact solutions of GR have received new clothes through their use in
string theory  and related higher-dimensional scenarios. For example,
the Taub-NUT solution, whose interpretation in conventional GR is
obscure, has undergone  a transformation (via Euclidean rotation, and
the addition of extra dimensions) into a solution to 11D supergravity,
which provides the lift to M-theory of a D6-brane.\cite{Emparan_d6} Another
interesting example is the use of the C-metric,
to describe black hole pair creation and exact solutions
for black holes on branes.\cite{Emparan_bhwall}
Much work extends the techniques of GR to cover Kaluza-Klein reductions, couplings to
higher-dimensional p-form gauge fields etc, but it would be desirable to
unravel the structure of the simplest situation -- {\it pure}
gravity ($R_{\mu\nu}=0$) in dimensions $d>4$. Along this line,
some solutions with qualitatively new features, such as
higher-dimensional rotating black holes,\cite{Emparan_mp} and the
instabilities of black branes,\cite{Emparan_gl} have been discovered.
A more systematic approach to the construction and study of such
solutions appears is still lacking.
\\
\\
\textbf{Cosmological solutions: selected themes} ({\it M. A. H. MacCallum})
\\
In this last talk of the workshop, MacCallum first discussed where
exact solutions impinge on the cosmological theory. The "standard
model" nowadays is the Friedmann-Lemaitre-Robertson-Walker (FLRW) model
with $k=0$, with cosmological constant $\Lambda$, and with inflation. Its
predictions fit the cosmic microwave background and the large-scale
structure, but worries remain: (i) because of the large number of
``initial'' assumptions, such as what is the inflaton, what is the
initial quantum state etc; (ii) there are uncertainties in its fit to
galactic structures; (iii) no convincing reasons are available for the
value of $\Lambda$.
     The role of exact cosmological solutions is that they capture
full non-linearity of the Einstein equations and provide tests of the
claims of the standard model. For example, phase-plane arguments
show that there are anisotropic solutions staying close to the FLRW models
for as long as one wishes.\cite{MacCallum02} Further, there are models
from which it is evident that inflation does not necessarily remove
anisotropy. It is not clear how generic these models are, not even what
``generic'' means. Exact solutions may be useful for observational data
analysis. For example, number counts can be explained by the
inhomogeneous Lemaitre-Tolman-Bondi models without an evolution of the
sources equally well as by the standard FLRW models with sources evolving.
By using exact cosmological solutions one can study (past and
future) asymptotics, gravitational waves in a non-linear regime (solutions
with $G_2$ symmetry groups) and see, within a more general framework, what
is special about the FLRW models.

    In the second part, MacCallum mentioned some progress made in the last
decade or so in the field of exact cosmological models. Most of the
results are summarized in Refs. [\cite{MacCallum02,MacCallum03}] where a number
of other citations can be found (for the references on vacuum cosmological
models, see also Ref. [\cite{JeFest}]).
New fluid solutions with a hypersurface homogeneity were found,
mainly as fixed points of dynamical systems. Other solutions were obtained
by generating mechanisms (in particular those with $p=\rho$), by separation
Ans\"atze, or by use of homotheties. More attention has been paid to
solutions with $\Lambda$. Methods using dynamical systems techniques,
symmetries of Hamiltonians, etc are summarized in Ref. [\cite{MacCallum02}]
(see also the contribution of Uggla to the workshop A.3 in these
Proceedings). Current efforts are extending to tilted models and cases
where self-similar asymptotics does not work. Interesting open questions
arise as to whether these new models can be used to test averaging,
or to study structure formation.

      Finally, MacCallum mentioned the problem of ``local''
objects in cosmology, in particular black holes in spacetimes
which are asymptotically FLRW. A simple example is given by
joining the Schwarzschild solution to a Whittaker solution in the
Einstein universe.\cite{MacCallum04}

\vskip 2mm
Most of the papers quoted bellow are also contained on the
qr-qc archives. Here we give the archive number only if the
journal citation is not available.
\def\PRD{{\it Phys. Rev. \rm D }}
\def\CQG{{\it Class. Quant. Grav. }}
\def\GRG{{\it Gen. Rel. Grav. }}

\end{document}